\documentstyle[12pt]{article}

\def\bq{\begin{eqnarray}}
\def\eq{\end{eqnarray}}
\def\g{\gamma}
\def\ra{\rightarrow}

\def\n{\nonumber}

\begin{document}

\begin{flushright}
April 1998
\end{flushright}
\vspace{1cm}
\begin{center}
{\bf $Z_N$-symmetry in gauge theories at finite temperatures. }\\
\vspace{1cm}
V.M. Belyaev\\
\vspace{1cm}
{\it ITEP, B.Cheremushkinskaya 25, 117259, Moscow, Russia}
\end{center}
\vspace{1cm}
\begin{abstract}
The role of $Z_N$ symmetry in  gauge theories
at finite temperatures is discussed.
This symmetry is studied
in terms of $A_0$-effective potential.
We consider  two- and four-dimensional
models
where the question on physical interpretation 
of minima of $A_0$-effective potential  can be considered 
exactly.
It is shown that 
the correct result for the  partition function 
can be obtained by summation of contributions near all minima
of the $A_0$-effective potential.

\vspace{0.5cm}

\noindent PACS number(s): 11.10.Wx; 11.15.-q
\end{abstract}
\newpage

\section{Introduction}
The Matsubara formalism
is a convenient tool to study static properties of
a theory at finite temperatures ~\cite{tem}.
A quantum field theory partition function  
at finite temperatures can be represented in the following form
\bq
Z=\int {\cal D}\phi_i(t,x)e^{-\int_0^\beta dt\int d^3x L_{eucl.}(\phi(t,x)_i};
\label{1}
\eq
where $L_{eucl.}$ is a Lagrangian in Euclidean space-time;
the (anti)periodic boundary conditions are imposed on the 
(fermi) boson
 quantum fields:
$\phi(t=0,x)=\phi(t=\beta,x)$; $\beta$ is inverse temperature: 
$\beta=1/T$.

It is important to note that due to the boundary conditions
it is not possible to apply the gauge $A_0=0$.
Indeed, let us consider a field configuration when
$A_0=const$ and $A_i$ are arbitrary gauge fields ($i=1,2,3$).
If we try to gauge off the constant gauge field $A_0$:
\bq
A_0^{'}=U(t)A_0U^\dagger(t)+
\frac{i}{g}U(t)\partial_0 U^{\dagger}(t)=0;
\n
\\
U(t)=e^{-ig tA_0};
\label{2}
\eq
then the boundary conditions for spatial components of gauge fields
will be changed:
\bq
A_i^{'}(t=\beta,x)=U(t=\beta)A_i(t=\beta,x)U^{\dagger}(t=\beta)
\n
\\
=
U(t=\beta)A_i(t=0,x)U^{\dagger}(t=\beta)
\n
\\
=
U(t=\beta)A_i^{'}(t=0,x)U^{\dagger}(t=\beta).
\label{3}
\eq
where $A_\mu=A_\mu^a t^a$, $t^a$ is a generator of a gauge group,
$g$ is a coupling constant.

It is possible to keep the periodic boundary
conditions only in the case when
$[t^a,U(t=\beta)]=0$ for any generator $t^a$ of a gauge
group. It means that the theory is invariant under the gauge
transformations $U^{\dagger}(t=0,\vec x)U(t=\beta,\vec x)$
 belonging to the center of the group.
The center of the $SU(N)$ group
is $Z_N$-discrete subgroup and the boundary conditions
are unchanged under  discrete shifts of gauge fields corresponding
to the gauge transformations from the center of the gauge group.
In the case of $SU(2)$ gauge group the shift is
$A_0^3\ra A_0^3+\pi T/g$.
It is important to note that the theory is not invariant
under a general shift of the gauge field which does not correspond
to the gauge transformation from the center of a gauge group.
It means that the effective potential for temporal component of the
gauge fields is not trivial when $T\neq 0$.
The effective potential $V_{eff.}(A_0)$ has a periodic structure
and  $Z_N$-symmetry
can be formulated in terms of the order parameter:
\bq
L(x)=\frac1{N_c}{\cal P} Tr\left( e^{ig\int_0^\beta A_0(t,x)dt}\right);
\label{4}
\eq
here ${\cal P}$ means P-ordering.

It was shown \cite{pot} that one- and two-loop effective
potentials have global minima where
\bq
<L(x)>=e^{i\frac{2\pi n}{N}}.
\label{5}
\eq
General proof of this statement has been made in \cite{pis}.

These minima are degenerated and correspond to the  different but 
physically equivalent
states of the {\bf Euclidean gauge theory} with fields in adjoint 
representation.
Fields in fundamental representation break  $Z_N$-symmetry
and $Z_N$-minima are not more degenerated.

It was a very attractive idea to interpretate such local 
minima of
the effective potential as  metastable
phases of a hot gauge theory and to use the decay of such
states in cosmology \cite{kaj}.
This interpretation of the local minima
was strongly criticized \cite{crit} and
it was pointed that this interpretation leads to a conclusion
that it may exist physical metastable states with
 negative entropy.
This remark is very important but
it is not enough to exclude the existence of the states where
the whole entropy is positive.
The absence of such "metastable states" has been demonstrated
in 2-dimensional Schwinger model \cite{smilga}.
 However this conclusion  may be a specific property of
2-dimensional systems.

So, it is clear that we have to find a clear physical
interpretation of such minima in the case of a simple
4-dimensional model.
Below we give a  physical explanation of
the so-called "metastable states" in the case
of $U(1)$ gauge theory at finite temperatures
with two type of charged particles.
The main conclusion which can be made from this
model is that the correct expression for partition
function
is a sum of partition functions near minima
of the effective potential. It means that
there are no negative entropy and negative specific heat
for these systems: they are just negative
corrections to the entropy and specific  heat.
It is also demonstrated that it is not correct
to consider $<L>$ as an order parameter which
confirms argumentation of \cite{smilga}.

At the end of the paper we consider $2-d$ Schwinger model
at finite temperatures.
It is demonstrated that in this model the effective
potential for the constant filed has a periodical structure
 in contrast to the effective potential for
variable gauge fields.
That is a property of 2-dimensional systems.

\section{Partition Function and Order Parameter}

Let us start our consideration from the formal
derivation of eq.(\ref{1}) in the case of $U(1)$ gauge theory.
For simplicity we consider a compact 3-dimensional space.
The partition function of a gauge theory 
in the gauge $A_0=0$ has the following form:
\bq
Z=\sum_n<n |e^{-\hat{H}t}|n >
\n
\\
\sim
\int {\cal D}A_i{\cal D}E_i {\cal D}\alpha e^{-\int_0^\beta dt\int d^3x
\left(H+iE_i\partial_0 A_i+i\alpha(\partial_i E_i-e\rho)-ieA_i j_i\right)}
\label{1a}
\eq
where $H=\frac12 (E^2+B^2)$ is a Hamiltonian; $E_i=\partial_0 A_i$ 
are electric fields;
$B_i=\frac12\epsilon_{ijk}\partial_jA_k$ are the magnetic fields; 
$\rho$ is a charge density, $j_i$ is an electric current,
$e$ is an electric charge.

The partition function (\ref{1a}) has 
 the Gauss law constraint:
\bq
\int {\cal D}\alpha e^{i\int_0^\beta dt\int d^3x\alpha(\partial_i E_i-\rho)}
\sim \delta[\partial_i E_i-e\rho].
\label{2a}
\eq

After integration by parts in eq.(\ref{1a}) of the term with
$\partial_i E_i$ it is not difficult to show that
in the case of compact space the partition function has a form
of Euclidean gauge theory with periodic boundary conditions
imposed on the gauge fields:
\bq
Z=\int{\cal D}A_\mu e^{-\int_0^\beta dt\int d^3x
\left(\frac14 F_{\mu\nu}^2-ie A_\mu j_\mu\right)}
\label{2aa}
\eq
where $F_{\mu\nu}=\partial_\mu A_\nu-\partial_\nu A_\mu$ is
a strength tensor, and $A_0=\alpha$. This formula
can be obtained in the case of nonabelian gauge theory
with a replacement $\partial_\mu\ra D_\mu=\partial_\mu
-igA_\mu$.

One can represent $\alpha$-field in the following form:
\bq
\alpha(t,x)=\alpha_0(t)+\alpha_x(t,x)
\label{3a}
\eq
where
fields $\alpha_0$ and $\alpha_x$ belongs to different orthogonal 
functional subspaces. It means that any field configuration 
$\alpha_x(t,x)$ is orthogonal to an arbitrary function
$\alpha_0$ and:
\bq
\int d^3x\alpha_x(t,x)=0
\label{4a}
\eq
The integration in the $\alpha_0$-subspace leads us to a conclusion (in the case
of compact  $x$-space case) that
\bq
\int {\cal D}\alpha_0  e^{i\int_0^\beta dt\int d^3x\alpha_0(\partial_i E_i-e\rho)}
\n
\\
=\int {\cal D}\alpha_0  e^{-ie\int_0^\beta dt\int d^3x\alpha_0(t)\rho(x,t)}
\sim \delta[e\int d^3x\rho(x,t)]
\label{5a}
\eq
which means that only {\bf neutral states} gives nonzero 
contribution.

Let us consider the order parameter (\ref{4})
in the case of $U(1)$ theory with light electrons ($T\gg m$) 
where $m$ is a mass
of  particles with charge $e$.

Taking into account the 
$\delta$-function (\ref{5a}) we can conclude that
the average value for the order parameter with a charge $e/2$ is
equal to zero:
\bq
<L(A_0(\vec x)>=<{\cal P}\exp 
\left(i\frac{e}{2}\int_0^\beta A_0(t,\vec x)dt\right)>=0,
\label{7a}
\eq
because of the fact that the presence of the Wilson line in the path
integral
is equivalent to consideration of a system with one
 heavy particle with a charge $1/2$ and it is not
possible to create a neutral system with one  particle
with a charge $1/2$ plus any large but finite number of particle with 
charge $1$.

In the imaginary time formalism
one can reproduce this result only in the case if we make $A_0$-integration
in the whole functional subspace $\alpha_0$. 
In high temperature limit we can apply saddle point approximation
near the points where
 $<L>=\pm 1$:
\bq
<L>=Z(L)/Z
=\frac{Z(L)_{A_0=0}+Z(L)_{A_0=2\pi T/e}}
{Z_{A_0=0}+Z_{A_0=2\pi T/e}}=0.
\label{8a}
\eq
Here
\bq
Z(L)=\int{\cal D}A_\mu L(A_0) \exp \left(-\int_0^\beta 
dt\int d^3x\frac14 F_{\mu\nu}^2\right)
\label{8aa}
\eq
Due to the $Z_2$ symmetry (in terms of the
order parameter $L$ with semi-integer charge) this result is exact.
 The relation (\ref{8a}) does not mean that we have a  confinement but 
it
is just
the consequence of the neutrality condition $Q=\int d^3x\rho(x)=0$.

So, it is clear that it is not correct to consider the value
$<L>$ as an order parameter. It confirms argumentation of
\cite{smilga} that instead of $<L>$ the asymptotic of the correlator
\bq
<L(\vec x)L^{\dagger}(0)>_{\vec x\ra\infty}
\label{8aaa}
\eq
can be considered as a correct order parameter of a theory.

Note that to obtain the correct result $<L>=0$ for the
Wilson line with an arbitrary fractional charge one
have to sum up over all minima of the effective potential.
Only in this case it is possible to obtain the correct
result for the partition function.

It is easy to construct a model with so-called "metastable state".
It is $U(1)$ model with two types of charged particles:
the first particles have a mass $m\ll T$ and integer charge $e$
and the second ones have a mass $M\gg T$ and semi-integer charge
$e/2$.
 
The order parameter $<L>$ with semi-integer charge
is not equal to zero because of the presence
of the massive particles with charge $e/2$.
The neutrality condition in this case means that
only states with odd numbers of particles with charge
$e/2$ give nonzero contribution into the partition function.
It is clear that the order parameter $<L>$ will be suppressed by a factor
$\sim e^{-m/T}$

It is possible to reproduce this result in the imaginary time formalism:
\bq
<L> &=& \frac{Z(L)_{A_0=0}+Z(L)_{A_0=2\pi T/e}}
{Z_{A_0=0}+Z_{A_0=2\pi T/e}}
\n
\\
&\simeq & \frac{Z_{A_0=0}-Z_{A_0=2\pi T/e}}
{Z_{A_0=0}+Z_{A_0=2\pi T/e}}
\n
\\
&\simeq &
V\int\frac{d^3k}{(2\pi)^3}e^{-\sqrt{k^2+M^2}/T};\;\; 
\label{10a}
\eq
where $V$ is 3-dimensional volume of a system.
This result is exact in the limit when
$\int\frac{d^3k}{(2\pi)^3}e^{-\sqrt{k^2+M^2}/T}\ll V^{-1}$.
Nonzero result for in (\ref{10a}) appears due to the difference 
for a free energy at $A_0=0$
and $A_0=\pi/(e \beta)$.
It is important to note that the correct result (\ref{10a})
for the partition function 
can be obtained only after integration over the all
values of $A_0$.

According to the neutrality condition only
the states with even numbers of heavy particles give
nonzero contribution to the partition function.
It is possible to check that the total contribution of
heavy particles into the partition function in present
 model has the following structure:
\bq
Z=\frac12[Z(0)+Z(1)+Z(2)+..]+[Z(0)-Z(1)+Z(2)+...]
\n
\\
=Z(0)+Z(2)+...
\label{11a}
\eq
where $Z(n)$ is a partition function of the system with
$n$-heavy particles, and the first term of eq.(\ref{11a}) in square
brackets corresponds to $A_0=0$ minimum of the effective
potential and the second one to the local minimum
of the potential where $<L>=-1$.

Note that one of the term of eq.(\ref{11a}) looks like
a partition function of particles with a wrong statistic.
The same wrong statistic in distribution functions appears
in the case of nonabelian theory with matter
in fundamental representation \cite{crit}.
So it becomes clear that the physical sense of this
local minimum: it is not a metastable state with such wrong statistic
but only the correction
to the  partition function.

However there is a possibility to interpritate these
states as a metastable phases of a theory in the case
when the compact coordinate is chosen in a spatial direction.
In this case such domain walls can be considered as physical
objects because in this case we do not consider a sum over all physical
states which is essential in the case of a hot gauge theory.

\section{Schwinger Model}

In Schwinger Model we are able to determine the thermodynamical 
properties  exactly.

The partition function has the following form:
\bq
Z=\int{\cal D}A_\mu{\cal D}\psi{\cal D}\bar{\psi}
e^{-\int_0^\beta dt\int dx\left(\frac14 F_{\mu\nu}^2+i\bar\psi\hat{D}\psi\right)};
\label{1b}
\eq
where
$\hat{D}=\g_\mu (\partial_\mu-ie A_\mu)$.

The simplest way to find the effective action for gauge fields $A_\mu$
in this model  is
to apply the local axial transformation:
\bq
\psi(t,x)\ra e^{i\g_3\phi(t,x)}\psi(t,x)
\n
\\
\bar\psi(t,x)\ra \bar\psi(t,x)e^{i\g_3 \phi(t,x)}
\label{2b}
\eq
The  determinant of fermions after the transformation
(\ref{2b}) have the following form:
\bq
\det {i\hat D}_{A_\mu}\ra
\det {i\hat D}_{A_{\mu}^{'}}
\label{3b}
\eq
where
\bq
A_\mu^{'}=A_\mu+\frac{i}{e}\varepsilon_{\mu\nu}\partial_\nu\phi(t,x).
\label{4b}
\eq
The strength tensor of the $A_\mu^{'}$ field is
\bq
F_{\mu\nu}^{'}=F_{\mu\nu}-\frac{i}{e}\partial^2\phi(t,x)
\label{5b}
\eq
So, it is clear that if we choose that
\bq
\phi(t,x)=\frac{i}{-\partial^2}F
\label{6b}
\eq
where $F=\frac12 \varepsilon_{\mu\nu}F_{\mu\nu}$
then the determinant will be equal to
the determinant of free fermions in the presence
of a constant gauge fields with zero strength tensor.

It means that the determinant is
\bq
\det (i\hat D)_{A_\mu}=\det (i\hat{D})_{A_\mu=const} J
\label{7b}
\eq
where $J$ is the Jacobian of the axial transformation
(\ref{2b}) (anomaly):
\bq
J=e^{i \frac{e}{\pi}\int_0^\beta dt\int dx F\phi}
\label{8b}
\eq

It is important to remark that the constant mode of the 
$\phi$ is excluded. 

It was noted in the Introduction that we can not gauge away the
constant gauge field. Thus, after integration over fermion fields 
the partition function (\ref{1b})
have the following form:
\bq
Z= Z_F(A_0=const)\int{\cal D}A_\mu
e^{-\int_0^\beta dt\int dx \left(\frac 14 F_{\mu\nu}^2+\frac{e^2}{\pi} F\frac1{-\partial^2}
F\right)}
\label{9b}
\eq
where $Z_F$ is a partition function for fermion field in the presence
of constant gauge field with $F=0$.

It is possible to rewrite the eq.(\ref{9b}) in more suitable form by
adding the auxiliary scalar field $\phi$:
\bq
Z= Z_F(A_0=const)Z^{-1}_B(m=0,\phi\ne const)\int{\cal D}A_\mu{\cal D}\phi
e^{-\int_0^\beta dt\int dx \left(\frac 14 F_{\mu\nu}^2+i\frac{e}{\pi} F\phi
+\frac12 (\partial_\mu \phi^2)\right)};
\label{10b}
\eq
where $Z_B$ is a partition function for a free massless boson field
and the constant mode of  field $\phi$ is excluded.
Otherwise, the integration over this constant mode leads us to a conclusion
that the topological charge of the system is equal to zero:
\bq
\int_0^\beta dt\int dx F=0
\label{10bb}
\eq

Now let us go back to the eq.(\ref{9b}).
Let us try to find the effective potential for the field $A_0(x)$.
The effective potential for the constant part of this field
has a periodic structure and comes from $Z_F(A_0)$ only:
\bq
V_{eff}(A_0=const)=\frac{e^2}{\pi} (A_0^2)_{|mod\;\; 2\pi T/e}
\label{11b}
\eq
At the same time the exact result for the effective potential of the nonzero modes
of the field $A_0(x)$ has no a periodic structure and can be obtained from
effective Lagrangian:
\bq
V_{eff}(A_0\neq const)=\frac{e^2}{\pi} A_0^2
\label{12b}
\eq

It is a specific property of $2-d$ theory.
In the case of $4-d$ theory this factorization property is absent.

\section{Conclusions}

The main question considered in present paper is the physical sense of so-called
"metastable states" in hot gauge theories.
It was shown that in $4-d$ QED such local minima of the effective
potential can not be treated as a physical states.
The only sense of such minima is a way to remove charged states from
the consideration.
It was shown that in the case of the simplest $U(1)$ theory
this conclusion can be proved exactly.
It was demonstrated that we have to integrate over all values
of static $A_0$ fields to obtain a physical result.
It has been shown that the mean value of Polyakov operator can not
be considered as an order parameter. 
It confirms an argumentation of Ref.\cite{smilga}.

We have to emphasize that a simple naive picture of
the confinement-deconfinement transition in terms of transitions
of $A_0$ gauge field from one minima to another is not correct.
The counter-example is the $U(1)$ $4-d$ model where it was
shown
that at low temperatures when these transitions are not suppressed and
the "order parameter" is equal to zero. This zero
value for $<L>$ means just the absence
of charged states at low temperatures and can not be treated as a 
confinement.

In the case of the Schwinger model all results can be obtained exactly.
It was demonstrated that the periodic structure of the effective
potential is a property of a constant nondynamical mode of the
gauge field $A_0$ only. The effective potential for the nonconstant
gauge fields has no periodic structure. 

These two examples clearly show that it is not correct to consider
local minima of the effective potential as metastable phases
of theory at finite temperatures.
Nevertheless such a minima of effective potentials may
appear in  gauge theories with one compact space coordinate.
In this case these minima of the effective potential can be considered as
real metastable states.
 
\section{Acknowledgements}

The author thanks J.-P. Blaizot and A.V. Smilga for stimulating discussions.
This research was sponsored in part by the INTAS Grant 93-0283,
CRDF Grant RP-2-132, and Swiss Grant 7SUPJ048716.

\end{document}